\documentclass[12pt]{iopart}
\usepackage{iopams}
\expandafter\let\csname equation*\endcsname\relax
\expandafter\let\csname endequation*\endcsname\relax
\usepackage{amsmath}

\usepackage{cite}
\usepackage{bm}
\usepackage{float}
\usepackage{mathtools}
\usepackage{multirow}

\usepackage{bbold}
\usepackage{bbm}
\usepackage{braket}

\usepackage{amsthm}
\usepackage{mathrsfs}
\usepackage[usenames]{xcolor}
\usepackage{gensymb}
\usepackage{appendix}
\usepackage{graphicx}
\usepackage[caption=false]{subfig}
\usepackage[normalem]{ulem}
\usepackage{color}

\usepackage{soul}
\usepackage{hyperref}

\usepackage{cancel}

\def\beq{\begin{equation}}
\def\eeq{\end{equation}}
\def\bea{\begin{eqnarray}}
\def\eea{\end{eqnarray}}

\begin{document}

\title{Nonlinear dynamics of the dissipative anisotropic two-photon Dicke model}

\author{Jiahui Li$^{1}$, Rosario Fazio$^{2,3}$ and Stefano Chesi$^{1,2,4,*}$}
\address{$^1$Beijing Computational Science Research Center, Beijing 100193, People's Republic of China}
\address{$^2$The Abdus Salam International Center for Theoretical Physics (ICTP), Strada Costiera 11, 34151 Trieste, Italy}
\address{$^3$Dipartimento di Fisica, Universit\`a di Napoli ``Federico II'', Monte S. Angelo, I-80126 Napoli, Italy}
\address{$^4$Department of Physics, Beijing Normal University, Beijing 100875, People's Republic of China}
\address{$^*$Author to whom any correspondence should be addressed.}

\ead{stefano.chesi@csrc.ac.cn}

\begin{abstract}
We study the semiclassical limit of the anisotropic two-photon Dicke model with a dissipative bosonic field and describe its rich nonlinear dynamics. Besides normal and `superradiant'-like phases, the presence of localized fixed points reflects the spectral collapse of the closed-system Hamiltonian. Through Hopf bifurcations of superradiant and normal fixed points, limit cycles are formed in certain regions of parameters. We also identify a pole-flip transition induced by anisotropy and a region of chaotic dynamics, which appears from a cascade of period-doubling bifurcations. In the chaotic region, collision and fragmentation of symmetric attractors take place. Throughout the phase diagram we find several examples of phase coexistence, leading to the segmentation of phase space into distinct basins of attraction.
\end{abstract}

\maketitle

\section{ Introduction }

The quantum Dicke model~\cite{1954Coherence}, describing a collective interaction of $N$ two-level systems (qubits or atoms) with a single bosonic mode, is an important paradigm for light-matter interactions. Besides realizing a second-order transition to a superradiant phase~\cite{Klaus1973Equilibrium,dimer2007proposed}, the Dicke model is applicable to a wide range of scenarios relevant to quantum technology~\cite{bishop1996application,ashhab2010qubit,beaudoin2011dissipation,puebla2016robust,lv2018quantum,zhang2018generalized,baksic2014controlling,zou2014implementation,joshi2016quantum,dalla2016dicke,larson2017some,fitzpatrick2017observation,gelhausen2017many,gelhausen2018dissipative} (see also Ref.~\cite{kirton2019introduction} for a recent review). In particular, various equilibrium and non-equilibrium phenomena, such as ground and excited state phase transitions~\cite{Hwang2015Universal,Puebla2016excited}, dissipative phase transitions~\cite{kessler2012dissipative,hwang2018dissipative}, and dynamical phase transitions~\cite{klinder2015dynamical,puebla2020finite,heyl2018dynamical}, have been investigated in the isotropic and anisotropic quantum Dicke model, as well as in other related systems~\cite{bishop2001time,xie2014anisotropic,liu2017universal,zhang2017analytical,2017Suppressing,wang2018quantum,cui2019two,2019Multiple,chen2012delocalized,shammah2017superradiance}.

An interesting variation of the Dicke model considers a collective two-photon coupling and is motivated by recent progress in achieving strong and ultra-strong light-matter interactions. In these regimes, multi-photon interaction processes (previously suppressed by the weak coupling strength) become more prominent~\cite{travvenec2012solvability,puebla2017protected,cheng2018nonlinear,felicetti2018two,maldonado2019squeezed,PhysRevA.104.053718,Minganti_2021}. Correspondingly, the two-photon Dicke model has gained considerable interest. Several studies highlight the collapse of discrete energy levels into a continuous band, occurring at a certain threshold coupling~\cite{felicetti2015spectral, duan2016two,cong2019polaron,rico2020spectral}. Furthermore, before the spectral collapse takes place, there is a phase transition to a `superradiant'-like phase, where the fourfold discrete symmetry is spontaneously broken~\cite{garbe2017superradiant,chen2018finite,xie2019generalized,cui2020nonlinear,garbe2020dissipation,ying2020quantum}. On the other hand, while stationary states of this model are relatively well understood, its dissipative nonlinear dynamics has not been investigated in detail.

Thanks to the collective nature of the coupling, Dicke models can be studied through the semiclassical approximation, valid in the limit of large $N$. For a regular single-photon interaction, nonlinear dynamics induced by counter-rotating terms leads to classical chaos in the strong-coupling regime. Away from the thermodynamic limit, the system undergoes a transition from quasi-integrability to quantum chaos, caused by the precursors of the quantum phase transition~\cite{emary2003chaos,emary2003quantum,hou2004decoherence,song2009spin,altland2012quantum,bastarrachea2015chaos,chavez2016classical,lobez2016entropy,ray2016quantum}. Significant efforts have been devoted to studying chaos in semiclassical and quantum regimes of closed Rabi and Dicke models~\cite{zhu2019single,wang2019effect,patra2019driven,patra2019chaotic,lerma2019dynamical}. Furthermore, including the effect of dissipative channels, nonlinear dynamics and chaotic behavior in a driven-dissipative setting \cite{kirton2018superradiant} and considering anisotropic couplings \cite{stitely2020nonlinear} have been recently discussed. The rich nonlinear dynamics of the one-photon Dicke model suggests that similar interesting phenomena can be found in the two-photon model as well.

With these motivations in mind, we present in this article a study of nonlinear dynamics in the two-photon Dicke model, including the effects of bosonic field dissipation and anisotropic couplings (i.e., where the rotating and counter-rotating terms are unbalanced). After introducing the model in Sec.~\ref{sec_model}, together with a detailed justification of the mean-field approximation (see \ref{app1}), we discuss three main dynamical behaviors allowed by the system: (i) Various types of stable fixed-point, see Sec.~\ref{sec:fixed_points}; (ii) Limit cycles arising from Hopf bifurcations of normal and superradiant fixed points, see Sec.~\ref{sec_lc}; (iii) Chaotic motion, see Sec.~\ref{sec_chaos}. The interplay of these dynamical regimes determines a complex behavior in parameter space, summarized by various phase diagrams. In particular, chaos emerges beyond a pole-flip transition point from a cascade of period-doubling bifurcations. We also find several types of phase coexistence, leading to fragmentation of phase space and sensitive dependence of the asymptotic dynamics on the initial state. The coexistence between different types of stable fixed points is discussed in more detail in ~\ref{sec_multi}. Finally, Sec.~\ref{sec_con} contains our concluding remarks. 

%%%%%%%%%%%%%%%%%%%%%%%%%%%%%%%%%%%%%%%%%%%%%%%%%%%%%%%%%%%%%%%%%%%%%%%%
\section{The model}\label{sec_model}
%%%%%%%%%%%%%%%%%%%%%%%%%%%%%%%%%%%%%%%%%%%%%%%%%%%%%%%%%%%%%%%%%%%%%%%%%%

The Hamiltonian of the two-photon anisotropic Dicke model can be expressed as follows (setting $\hbar=1$)
\begin{align}
\label{eqH1}
\hat{H} =\omega_0\hat{a}^{\dagger}\hat{a}+\frac{\omega_q}{2}\sum_{j=1}^N\hat{\sigma}_z^{(j)} +\frac{g}{N}\sum_{j=1}^N[(\hat{a}^2\hat{\sigma}_+^{(j)}+\hat{a}^{\dagger2}\hat{\sigma}_-^{(j)})
 +\lambda(\hat{a}^2\hat{\sigma}_-^{(j)}+\hat{a}^{\dagger2}\hat{\sigma}_+^{(j)})],
\end{align}
where $\hat{a}$ ($\hat{a}^{\dagger}$) is the annihilation (creation) operator of the bosonic field with frequency $\omega_0$, $\vec{\hat{\sigma}}^{(j)}$ are the Pauli operators of two-level system $j$ (qubit or atom), while $\hat{\sigma}_{\pm}^{(j)}=\frac{1}{2}(\hat{\sigma}_{x}^{(j)}\pm i\hat{\sigma}_{y}^{(j)})$. The identical qubits have energy transition frequency $\omega_q$ and two-photon interaction strength $g$ with the bosonic field. The parameter $\lambda$ (we assume $\lambda>0$) models an imbalance of rotating and counter-rotating couplings, which may be altered by the intensity of the electric and magnetic fields in circuit QED implementations~\cite{felicetti2018two} or the power of lasers in trapped ions setups~\cite{puebla2017protected,cheng2018nonlinear}. Differently from the one-photon case, the two-photon Dicke model features a four-fold symmetry with the generalized parity operator $\Pi=(-1)^N\bigotimes_{j=1}^{N}\hat{\sigma}_z^{(j)}e^{i\pi\hat{a}^{\dagger}\hat{a}/2}$~\cite{felicetti2015spectral,chen2018finite}. The Hamiltonian is invariant under the parity transformation ($\hat{a}\rightarrow i\hat{a}, \hat{\sigma}_{x,y}\rightarrow -\hat{\sigma}_{x,y}$). In terms of collective angular momentum operators $\vec{\hat{J}}=\frac{1}{2}\sum_{j=1}^{N}\vec{\hat{\sigma}}^{(i)}$, we rewrite the Hamiltonian as:
\begin{align}
\label{eqH2}
\hat{H}=\omega_0\hat{a}^{\dagger}\hat{a}+\omega_q\hat{J}_z
+\frac{g}{N}[(1+\lambda)\hat{X}\hat{J}_x+(1-\lambda)\hat{Y}\hat{J}_y],
\end{align}
where $\hat{X}=\hat{a}^2+\hat{a}^{\dagger2}$ and $\hat{Y}=i(\hat{a}^2-\hat{a}^{\dagger2})$. Previous studies have shown that the two-photon Dicke model admits a `superradiant'-like phase transition, after which the collective pseudospin acquires a macroscopic mean value. At the same time, the bosonic field is driven to a squeezed vacuum state. While the expectation value of $\hat{a}$ remains zero, squeezed quantum fluctuations lead to a phase with non-zero photon number, which is dubbed `superradiant'~\cite{garbe2017superradiant}. Here we consider this model in an open environment by including a dissipation channel for the bosonic field, with decay rate $\kappa$. The system evolution is described by a standard Lindblad master equation:
\begin{align}
\label{eqME}
\dot{\hat{\rho}}=-i[\hat{H},\hat{\rho}]+\kappa(2\hat{a}\hat{\rho} \hat{a}^{\dagger}-\hat{a}^{\dagger}\hat{a}\hat{\rho}-\hat{\rho} \hat{a}^{\dagger}\hat{a}).
\end{align}
Note that the parity operator $\Pi$ is still a symmetry of the system, i.e., $\Pi^\dag \hat{\rho} \Pi$ is also a solution of Eq.~(\ref{eqME}). We will be mainly concerned with the large-$N$ limit when spins can acquire a macroscopic population at coupling strengths comparable to the bosonic field frequency. This justifies a standard mean-field approximation, i.e., we decouple bosonic-qubit correlations as $\langle \hat{B}\hat{Q}\rangle\simeq \langle \hat{B}\rangle\langle \hat{Q} \rangle$, obtaining the following set of nonlinear equations:
\begin{align}
&\frac{d\langle \hat{X}\rangle }{dt}=-2\kappa \langle \hat{X} \rangle-2\omega_0 \langle \hat{Y} \rangle-2g(1-\lambda)\left(2\langle \hat{a}^{\dagger}\hat{a} \rangle+1\right)s_y, \label{eqMFfirst}\\
&\frac{d\langle \hat{Y} \rangle}{dt}=-2\kappa \langle \hat{Y} \rangle+2\omega_0 \langle \hat{X} \rangle+2g(1+\lambda)\left(2\langle \hat{a}^{\dagger}\hat{a} \rangle+1\right)s_x,  \\
&\frac{d \langle \hat{a}^{\dagger}\hat{a} \rangle}{dt}=-2\kappa \langle \hat{a}^{\dagger}\hat{a} \rangle +g(1+\lambda)\langle \hat{Y} \rangle s_x -g(1-\lambda)\langle \hat{X} \rangle s_y,  \\
&\frac{d s_x}{dt}=\frac{1}{N}\left[-\omega_zs_y+g(1-\lambda)\langle \hat{Y} \rangle s_z\right],   \\
&\frac{d s_y}{dt}=\frac{1}{N}\left[\omega_zs_x-g(1+\lambda)\langle \hat{X} \rangle s_z\right],   \\
\label{eqMFlast}
&\frac{d s_z}{dt}=\frac{1}{N}\left[g(1+\lambda)\langle \hat{X} \rangle s_y-g(1-\lambda)s_x\langle \hat{Y} \rangle\right].
\end{align}
In Eqs.~(\ref{eqMFfirst}--\ref{eqMFlast}) we rescaled the expectation values of the spin operators as $\vec{s}=\langle 2\vec{\hat{J}} \rangle/N$, and defined the collective spin frequency $\omega_z=N\omega_q$. As only the bosonic field dissipation is taken into consideration, we can restrict the collective spin evolution to the unit sphere, giving a five-dimensional phase space. The above equations contain explicitly $N$, meaning that system dynamics will be influenced by the number of qubits. The influence will be most obvious on the decoherence time and non-stable dynamics. On the other hand the steady-state is independent of $N$.

One interesting point about Eqs.~(\ref{eqMFfirst}--\ref{eqMFlast}) is that, at variance with $\vec{s}$, the bosonic variables are not scaled with $N$ and can remain small. In other words, the mean-field treatment is still valid for bosonic-field states which do not have a well-defined classical limit, as long as the atomic field has a macroscopic population. In \ref{app1} we show explicit comparisons to the exact evolution from Eq.~(\ref{eqME}), confirming the validity of Eqs.~(\ref{eqMFfirst}--\ref{eqMFlast}) in the large $N$ limit.

%%%%%%%%%%%%%%%%%%%%%%%%%%%%%%%%%%%%%%%%%%%%%%%%%%%%%%%
\section{Stable fixed points}\label{sec:fixed_points}
%%%%%%%%%%%%%%%%%%%%%%%%%%%%%%%%%%%%%%%%%%%%%%%%%%%%%%%

We first discuss the stationary states of the nonlinear dynamics, whose stability can be determined in a standard manner by the Jacobian matrix and Routh-Hurwitz criterion~\cite{strogatz2018nonlinear,dejesus1987routh}. Various stable fixed points appear in our system, whose acronyms are summarized in Table~\ref{table1}. We will also discuss different types of coexistence phases and limit cycles which, for easier reference, are also listed in Table~\ref{table1}. The simplest type of fixed points features zero photon number and trivial spin states:
\begin{align}
\label{eqNP}
{\rm{NP}_{\downarrow}}:\,\,\,&s_z=-1, \langle \hat{a}^{\dagger}\hat{a}\rangle=0, \notag \\
{\rm{NP}_{\uparrow}}:\,\,\, &s_z=1, \langle \hat{a}^{\dagger}\hat{a}\rangle=0.
\end{align}
The ${\rm{NP}_{\uparrow}}$ state is stable for
\begin{align}
\label{eqlamt}
\lambda > \lambda_t=\sqrt{\frac{4\kappa^2+(2\omega_0+\omega_q)^2}{4\kappa^2+(2\omega_0-\omega_q)^2}},
\end{align}
while ${\rm{NP}_{\downarrow}}$ is only stable for $\lambda < \lambda_t$. Therefore, the anisotropic parameter $\lambda$ can induce a dramatic change of atomic dynamics, where all down-spin states $|\downarrow\rangle$ transform to up-spin state $|\uparrow\rangle$ (a pole-flip transition~\cite{stitely2020nonlinear}). When $\lambda < \lambda_t$ there are two additional phase boundaries for the ${\rm{NP}_{\downarrow}}$ fixed point. The stability conditions are:
\begin{align}
\label{gt2}
& g \leq  g_{t2}=\sqrt{\frac{\omega_z}{\omega_0}}g_{t1}, \\
\label{gt3}
& g \geq g_{t3}=\sqrt{\frac{\omega_0\omega_z(\omega_0^2+\kappa^2)}{(1-\lambda^2)^2}} g_{t1}^{-1}, 
\end{align}
where we defined:
\begin{equation}
\label{gt1}
g_{t1}=\sqrt{\frac{\omega_0^2+\kappa^2}{1+\lambda^2+\sqrt{(2\lambda)^2-\kappa^2/\omega_0^2(1-\lambda^2)^2}}}.
\end{equation}  
In other words, ${\rm{NP}_{\downarrow}}$ has an instability window for the intermediate range of couplings $ g_{t2}< g <g_{t3} $. The phase boundary with the superradiant state of the closed model is recovered by setting $\kappa=0$ in the above expression of $g_{t2}$. Notice also that $g_{t3} \to \infty$ if $\lambda=1$, i.e., the upper critical line only exists for the anisotropic model.

%%%%%%%%%%%%%%%%%%%%%%%%%%%%%%%%%%%%%%%%%%%%%%%%%%%%%%%%%%%%%%%%%%%%

\begin{table} \footnotesize
\label{table1}
\centering
\setlength{\tabcolsep}{3mm}{
\begin{tabular}{|c|l|}
\hline
$\rm{NP}_{\downarrow}$ ($\rm{NP}_{\uparrow}$) & normal phase: all spins in $|\downarrow \rangle$ ($|\uparrow \rangle$) 
\\ \hline
$\rm{SP}$ & `superradiant'-like phase, with macroscopic occupation of the atomic field \\ \hline
$\rm{U}_0$ & localized phase: $\langle \hat{a}^\dag \hat{a}\rangle \to \infty $,  originating from a dissipative `spectral collapse' 
\\ \hline
$\rm{B}_\downarrow$ ($\rm{B}_\uparrow$) &  bistable phase: the stady-state, $\rm{SP}$ or $\rm{NP}_{\downarrow}$ ($\rm{NP}_{\uparrow}$), depends on initial conditions 
\\  \hline
$\rm{C}_\downarrow$ ($\rm{C}_\uparrow$) & $\rm{U}_0$ and $\rm{NP}_{\downarrow}$ ($\rm{NP}_{\uparrow}$) coexist in this phase, similarly to $\rm{B}_\downarrow$ ($\rm{B}_\uparrow$) 
\\  \hline
$\rm{LC}$ &  limit cycle  
\\ \hline
\end{tabular}}
\caption{ {Acronyms of different phases. $\rm{NP}_{\downarrow,\uparrow},\rm{SP},\rm{U}_0$ denote fixed points, discussed in Sec.~\ref{sec:fixed_points}. Periodic motion in the LC phase is presented in Sec.~\ref{sec_lc}.}}
\label{table1}
\end{table}
%%%%%%%%%%%%%%%%%%%%%%%%%%%%%%%%%%%%%%%%%%%%%%%%%%%%%%%%%%%%%%%%%%%%

By searching for nontrivial stationary conditions, the following `superradiant'-like fixed points are found:
\begin{align}
\label{eqSP}
{\rm{SP}}:\,\,\,&s_{z}=-\frac{\omega_0}{2\omega_z}+\sqrt{\left(\frac{\omega_0}{2\omega_z}\right)^2+1-\frac{g_{t1}^2}{g^2}}, \notag \\
&\langle \hat{a}^{\dagger}\hat{a}\rangle=-\frac{1}{2}\left(\frac{\omega_z}{\omega_0}\frac{ g_{t1}^2}{g^2}\frac{1}{s_{z}}+1\right),
\end{align}
in which atomic and bosonic fields are spontaneously occupied. By symmetry, a pair of states with opposite value of $(\pm\langle \hat{X}\rangle,\pm\langle\hat{Y} \rangle,\pm s_{x,y})$ appear in the SP phase. A SP fixed point requires:
\begin{align}
\label{lambda_range}
\frac{\sqrt{\omega_0^2+\kappa^2}-\omega_0}{\kappa}< \lambda < \frac{\sqrt{\omega_0^2+\kappa^2}+\omega_0}{\kappa}.
\end{align}
Beyond Eq.~(\ref{lambda_range}) only normal-phase solutions are possible, thus we will always consider values of $\lambda$ within this range. Furthermore, a physical SP solution only exists for:
\begin{equation}
\label{gt4}
g> g_{t4}=
\left \{
\begin{array}{ll}
g_{t2} & {\rm for}~\omega_z \leq  \omega_0/2, \\
\sqrt{\frac{4\omega_z^2}{4\omega_z^2+\omega_0^2}}g_{t1} & {\rm for}~\omega_z> \omega_0/2.
\end{array}
\right.
\end{equation}
As it turns out, for $\lambda\leq 1$ the SP fixed point is stable when $g_{t4}<g<g_{t1}$. For $\lambda>1$ the stability condition is more complex, and we can only compute it numerically. 

Finally, we recall that in the closed system the continuous phase transition to the superradiant phase is followed at larger $g$ by a `spectral collapse'~\cite{felicetti2015spectral,garbe2017superradiant}, as the coupling strength becomes comparable to the cavity frequency ($g_c = \omega_0/2$ if $\lambda=1$). At this collapse point, the discrete spectrum of $H$ is transformed into a continuous band. This dynamical feature has not been destroyed by dissipation: We see from Eq.~(\ref{eqSP}) that $s_z \to 0$ when $g\to g_{t1}$, leading to a divergent photon number. Therefore, at $g_{t1}$ the SP states evolve continuously to a pair of localized fixed points:
\begin{align}
\label{eqU0}
{\rm{U}_0}:\,\,\, s_{z,y}\rightarrow 0, \,\, s_{x}\rightarrow \pm1, \,\, \langle &\hat{a}^{\dagger}\hat{a}\rangle \rightarrow \infty.
\end{align} 
For $g>g_{t1}$, initial conditions within the basin of attraction of such localized fixed point lead to rapid growth of photon number, reminding the collapse of numerous energy levels occurring in the closed system. The spectral collapse point of the closed system is recovered by setting $\kappa=0$ in Eq.~(\ref{gt1}). This critical coupling is shifted by bosonic field dissipation to a larger value (e.g., $g_{t1} =\sqrt{\omega_0^2+\kappa^2}/2$ if $\lambda=1$).

%%%%%%%%%%%%%%%%%%%%%%%%%%%%%%%%%%%%%%%%%%%%%%%%%%%%%
\subsection{Phase diagram in the ($\omega_z,g$) plane}
%%%%%%%%%%%%%%%%%%%%%%%%%%%%%%%%%%%%%%%%%%%%%%%%%%%%%

\begin{figure}
	\centering
	\includegraphics[width=0.7\textwidth]{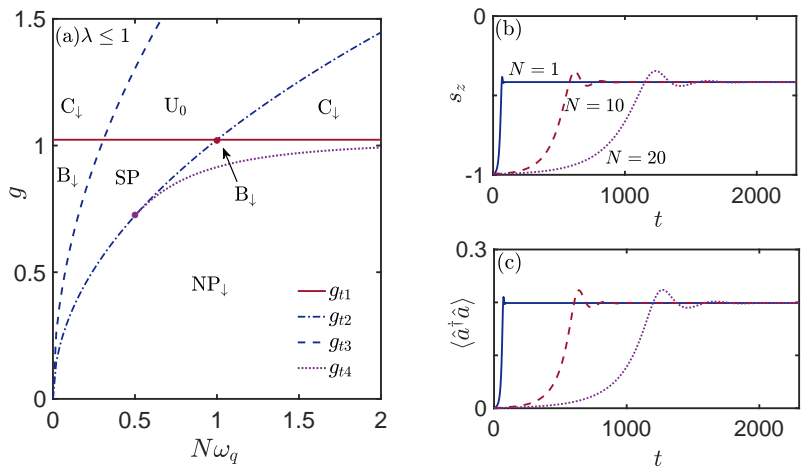}
	\caption{\label{fig1} Left panel: phase diagram for stable fixed points in the $(\omega_z,g)$ plane. This phase diagram is applicable for $\lambda \leq 1$. For $\lambda>1$, additional regions with limit cycles appear and a pole-flip transition occurs at $\lambda>\lambda_t$ (see Sec.~\ref{sec:phasediagram2}). The two dots mark the intersection of $g_{t2}$ with $g_{t4}$ and $g_{t1}$, located at $\omega_z/\omega_0=1/2$ and 1, respectively. Right panels: system evolution in $\rm{SP}$ for $g=0.6,\omega_z=0.2$, and different values of $N$. Other parameters are $\omega_0=\kappa=1$ and $\lambda=0.5$.}
\end{figure}

We now discuss explicit phase diagrams of stable fixed points in different parameter regimes. First, we take the qubit collective frequency $\omega_z$ and the coupling strength $g$ as variables, which for $\lambda<1$ leads to a phase diagram with the structure of Fig.~\ref{fig1}(a). Note that Eq.~(\ref{eqSP}) and all the expressions for the phase boundaries (except for $\lambda_t$) do not depend on $N$ explicitly if $\omega_z = N \omega_q$ is kept constant. Furthermore, for large $N$ also the boundary $\lambda_t$ approaches a well defined limit ($\lambda_t \to 1$). As shown in panels (b) and (c) of Fig.~\ref{fig1}, $N$ has an important effect on the timescale to reach the stationary state. However, the phase diagram of Fig.~\ref{fig1}(a) and all stationary states are independent of $N$. 

In Fig.~\ref{fig1}(a), a regular second-order phase transition between the $\rm{NP}_{\downarrow}$ and SP phases occurs at $\omega_z<\omega_0/2$, when $g_{t2}$ and $g_{t4}$ coincide. At $g=g_{t2}$ the normal state becomes unstable and `superradiant'-like solutions appear. This line is also identified as a pitchfork bifurcation point by bifurcation theory~\cite{strogatz2018nonlinear}. Examples of time evolution in the SP phase are shown in Fig.~\ref{fig1}(b) and (c).

The situation is different at $\omega_z>\omega_0/2$, when the phase boundaries $g_{t2}$ and $g_{t4}$ do not coincide. This determines a bistable region $g_{t4}< g < g_{t2}$. We denote with $\rm{B}_{\downarrow}$ the portion where $\rm{NP}_{\downarrow}$ and $\rm{SP}$ are both stable, i.e., if the additional constrain $g<g_{t1}$ is taken into account. As discussed already, the SP fixed point evolves to $\mathrm{U}_0$ at $g=g_{t1}$. Therefore, when $g_{t1}< g < g_{t2}$ the coexistence is between $\rm{NP}_{\downarrow}$ and $\rm{U}_0$, and is labeled as $\rm{C}_{\downarrow}$. In such coexistence regimes, the final state is determined by the initial condition as each stable point has its own basin of attraction. We refer to \ref{sec_multi} for a more detailed study of system evolution in the multi-stable regimes.

The last feature of the phase diagram in Fig.~\ref{fig1}(a) is the reentrant transition of the normal phase, which occurs for $g>g_{t3}$ and leads to a second pair of bistable regions $\rm{B}_{\downarrow},\rm{C}_{\downarrow}$ on the left side of the phase diagram. Since $g_{t3}|_{\lambda=1}\rightarrow\infty$, these $\rm{B}_{\downarrow},\rm{C}_{\downarrow}$ regions will shrink and eventually disappear when approaching the isotropic limit.

From the above discussion we see that, by increasing the coupling strength $g$ from zero to large values, the behavior of the system can be quite different depending on other system parameters. In particular, the critical lines $g_{t1}$ and $g_{t2}$ intersect at $\omega_z=\omega_0$. Therefore, in the range $\omega_0/2 <\omega_z < \omega_0$ the system enters the various phases as $ \rm{NP}_{\downarrow} \rightarrow \rm{B}_{\downarrow} \rightarrow \rm{SP} \rightarrow \rm{U}_0 \rightarrow \rm{C}_{\downarrow}$. Instead, for $\omega_z>\omega_0$ the sequence is modified to $ \rm{NP}_{\downarrow} \rightarrow \rm{B}_{\downarrow} \rightarrow \rm{C}_{\downarrow} \rightarrow \rm{U}_0 $. At variance with the one-photon Rabi model (where a large detuning $\omega_z/\omega_0$ is favorable to the formation of a superradiant phase), here the system does not support a `superradiant'-like state when $\omega_z\rightarrow \infty$. In this limit, the bistable region $\rm{B}_{\downarrow}$ shrinks to zero and a direct transition $\rm{NP}_{\downarrow} \rightarrow \rm{C}_{\downarrow}$ occurs. Such behavior would occur by considering $N \to \infty$ at fixed $\omega_q$.

%%%%%%%%%%%%%%%%%%%%%%%%%%%%%%%%%%%%%%%%%%%%%%%%%%%%%
%%%%%%%%%%%%%%%%%%%%%%%%%%%%%%%%%%%%%%%%%%%%%%%%%%%%%

%%%%%%%%%%%%%%%%%%%%%%%%%%%%%%%%%%%%%%%%%%%%%%%%%%%%%
\subsection{ Phase diagram in the ($\lambda,g$) plane }\label{sec:phasediagram2}
%%%%%%%%%%%%%%%%%%%%%%%%%%%%%%%%%%%%%%%%%%%%%%%%%%%%%

%%%%%%%%%%%%%%%%%%%%%%%%%%%%%%%%%%%%%%%%%%%%%%%%%%%%%
\begin{figure}
	\centering
	\includegraphics[width=0.65\textwidth]{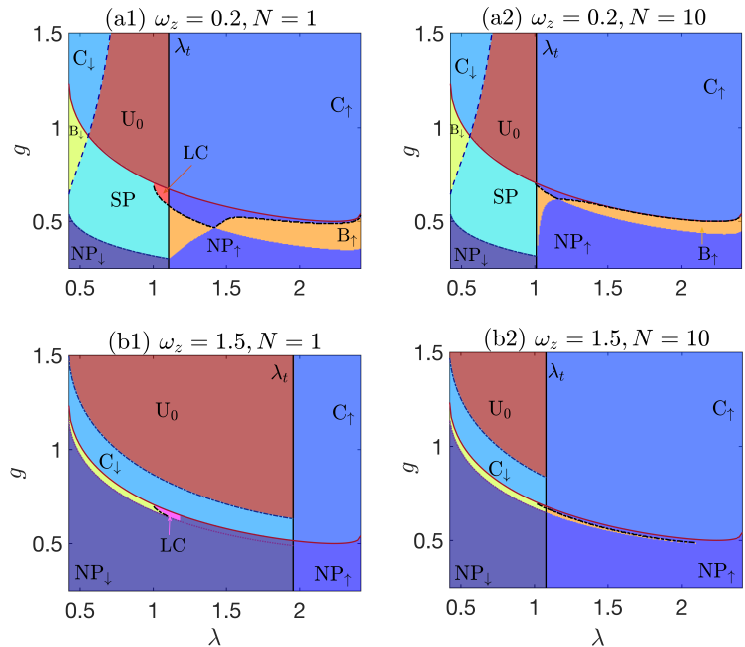}
	\caption{\label{fig2} Phase diagrams in the $(g,\lambda)$ plane. We use $N=1$ and $N=10$ for the left and right panels, respectively, while $\omega_z=0.2$ and $1.5$ (in units of $\omega_0$) in the upper and lower panels, respectively. The thick black vertical lines indicate the pole-flip transition point $\lambda_t$. The thick black dot-dashed lines indicate a Hopf bifurcation $g_t'$, at which stable oscillation phases $\rm{LC}$ appear (only shown for $\lambda<\lambda_t$). The solid red thin line is $g_{t1}$ while the other thin boundaries  at $\lambda < \lambda_t$ follows the same line style of Fig.~\ref{fig1}(a) for $g_{t2}$ (dot-dashed), $g_{t3}$ (dashed), and $g_{t4}$ (dotted).}
\end{figure}
%%%%%%%%%%%%%%%%%%%%%%%%%%%%%%%%%%%%%%%%%%%%%%%%%%%%%

To assess the influence of unbalanced rotating and counter-rotating couplings, we show in Fig.~\ref{fig2} phase diagrams of stable fixed points in the $(\lambda,g)$ plane. We also compare $N=1$ phase diagrams (left panels) to $N=10$ (right panels), representative of the large-$N$ limit. With respect to Fig.~\ref{fig1}(a), an obvious difference is the appearance of the fixed point $\rm{NP}_{\uparrow}$ in the regime $\lambda>\lambda_t$. The pole-flip transitions $\lambda_t$ are marked by vertical black lines in Fig.~\ref{fig2}, dividing each phase diagram into two main parts. For small to moderate $N$, the position of this pole-flip transition has a sensitive dependence on $\omega_z$, which can be seen comparing panels (a1) and (b1). The dependence is non-monotonic, with the maximum $\lambda_t$ occurring at $\omega_z = 2\omega_0$. On the other hand, as seen from panels (a2) and (b2), the dependence of $\omega_z$ is much weaker at large $N$, when $\lambda_t \simeq 1$. 

The $\lambda<\lambda_t$ region is quite complex, being determined by the critical lines $g_{t1},g_{t2}, g_{t3},g_{t4}$ discussed already. Therefore, the same phases of Fig.~\ref{fig1}(a) appear here. Like in Fig.~\ref{fig1}(a), the phase boundaries are independent of $N$ and there is a marked difference between the regimes $\omega_z<\omega_0/2$ (upper panels) and $\omega_z>\omega_0/2$ (lower panels). Instead, the region $\lambda>\lambda_t$ has a simpler structure dominated by a $\rm{NP}_{\uparrow}$ fixed point at $g<g_{t1}$ and a coexistence region ${\rm C}_\uparrow$ for $g> g_{t1}$. When $\lambda>\lambda_t$ the fixed point $\rm{NP}_{\uparrow}$ is always stable, thus SP only appears in a nontrivial bistable region ${\rm B}_\uparrow$ which shrinks to zero at large $N$.  As seen by a comparison of upper a lower panels of Fig.~\ref{fig2}, the ${\rm B}_\uparrow$ region also rapidly shrinks with $\omega_z$, confirming that a large $\omega_z$ is detrimental to the SP phase.

In concluding this section, we stress that the knowledge of stable fixed points is not sufficient to characterize the long-time dynamics. In panels (a1) and (b1) of Fig.~\ref{fig2} we also indicate the presence of limit cycles (LC), occurring for $1 \leq \lambda<\lambda_t$. These limit cycles originate from a Hopf bifurcation of the SP fixed point, which becomes unstable in these regions. Therefore, in panel (a1) the limit cycle does not coexist with any stable fixed point (orange-red region), while in panel (b1) we find the coexistence of the limit cycle and $\rm{NP}_{\downarrow}$ (magenta region). When $\lambda>\lambda_t$ the system dynamics is more involved: Besides stable fixed points and limit cycles, we find the occurrence of chaos. For the moment, in the regime $\lambda>\lambda_t$ we have only shown the phases corresponding to stable fixed points. A detailed analysis of limit cycles is presented in the following Sec.~\ref{sec_lc}, while the coexistence of stable fixed points, chaotic dynamics, and limit cycles at  $\lambda>\lambda_t$ is discussed in Sec.~\ref{sec_chaos}. A summary of various transitions encountered in parameter space (including chaos and limit cycles) is presented in Table~\ref{table2}.

%%%%%%%%%%%%%%%%%%%%%%%%%%%%%%%%%%%%%%%%%%%%%%%%%%%%%%%%%%%%%%%%%%%%
\begin{table} \footnotesize
\label{table2}
\centering
\setlength{\tabcolsep}{2mm}{
\begin{tabular}{|c|l|}
\hline
$\rm{NP}_{\downarrow}\rightarrow\rm{SP}$ & `superradiant'-like phase transition   
\\ \hline
$\rm{SP}\rightarrow\rm{U}_0$, ~ $\rm{B}_{\uparrow,\downarrow}\rightarrow \rm{C}_{\uparrow,\downarrow}$  & dissipative version of the SP `spectral collapse' (Sec.~\ref{sec:fixed_points})
\\ \hline
 $\rm{NP}_{\downarrow}\rightarrow\rm{NP}_{\uparrow}$  & pole-flip transition (Sec.~\ref{sec:phasediagram2})
\\ \hline
$\rm{NP}_{\downarrow}\rightarrow\rm{LC}_{NP_{\downarrow}}$,~ $\rm{SP}\rightarrow \rm{LC}_{\rm{SP}}$ & Hopf bifurcations from fixed points to limit cycles (Sec.~\ref{sec_lc}) 
\\ \hline
$\rm{LC}_{NP_{\downarrow}},\rm{LC}_{\rm{SP}} \rightarrow \rm{chaos}$ &
cascade of period-doubling bifurcations, leading to chaos (Sec.~\ref{sec_chaos})   
\\ \hline                 
\end{tabular}}
\caption{Summary of the various phase transitions discussed in the main text.}
\label{table2}
\end{table}
%%%%%%%%%%%%%%%%%%%%%%%%%%%%%%%%%%%%%%%%%%%%%%%%%%%%%%%%%%%%%%%%%%%%}

%%%%%%%%%%%%%%%%%%%%%%%%%%%%%%%%%%%%%%%%%%%%%%%%%%%%%
\section{Stable oscillations}\label{sec_lc}
%%%%%%%%%%%%%%%%%%%%%%%%%%%%%%%%%%%%%%%%%%%%%%%%%%%%%

%%%%%%%%%%%%%%%%%%%%%%%%%%%%%%%%%%%%%%%%%%%%%%%%%%%%%
\begin{figure}
	\centering
	\includegraphics[width=0.65\textwidth]{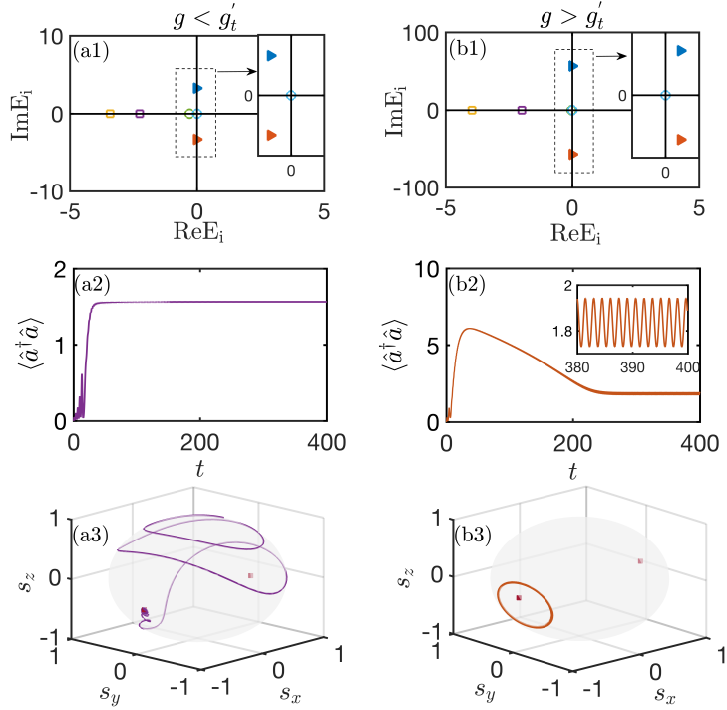}
	\caption{\label{fig3} Periodic orbit formed by $\rm{SP}$ via Hopf bifurcation. (a1) and (b1): Eigenvalues of the Jacobian matrix of $\rm{SP}$ before and after bifurcation. (a2) and (a3): Evolution of photon number and qubit trajectories before bifurcation. (b2) and (b3): Evolution after bifurcation. By setting $\omega_0=1$, we use $g=0.579$ for the left panels and $g=0.64$ for the right panels. The other parameters are $N=1$, $\lambda=1.2$, and $\omega_z=0.8$.}
\end{figure}
%%%%%%%%%%%%%%%%%%%%%%%%%%%%%%%%%%%%%%%%%%%%%%%%%%%%%

In non-linear dynamics, a Hopf bifurcation is a simple but important type of dynamic bifurcation. It describes the formation of a stable periodic orbit from a fixed point which, by varying system parameters, has lost its stability. In our system we find two Hopf bifurcations, marked by thick curves in Fig.\ref{fig2}. The first Hopf bifurcation, $g_t'$, is shown as a thick dot-dashed curve and originates from the SP fixed point. As shown in Fig.~\ref{fig3}, for $g<g_t'$ near the transition line the $\rm{SP}$ fixed point is stable: Panel (a1) shows that no eigenvalue of the associated Jacobian matrix has a positive real part. On the other hand, a pair of conjugate eigenvalues have crossed the imaginary axis in panel (b1), where $g>g_t'$. Before the bifurcation, as shown in panels (a2) and (a3), the system photon number and qubit trajectory converge to well-defined values, given by Eq.~(\ref{eqSP}). After the bifurcation, the photon number in panel (b2) shows persistent oscillations and the limit cycle becomes especially obvious from the spin trajectory on the Bloch sphere, shown in panel (b3). Here, for clarity, we only plot the system trajectory for relatively large times, $350<t<400$ (in units of $\omega_0^{-1}$), such that the system has already approached the stable periodic orbit. Due to symmetry, limit cycles bifurcating from $\rm{SP}$ appear in pairs. Only one of them is shown in Fig.~\ref{fig3}.

%%%%%%%%%%%%%%%%%%%%%%%%%%%%%%%%%%%%%%%%%%%%%%%%%%%%%
\begin{figure}
	\centering
	\includegraphics[width=0.65\textwidth]{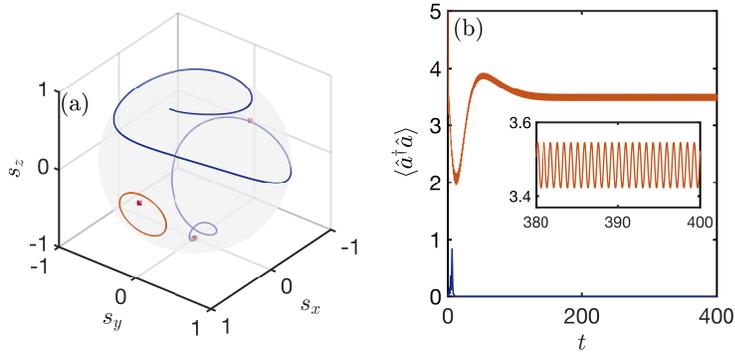}
	\caption{\label{fig4} Coexistence of a periodic orbit and the stable fixed point $\rm{NP}_{\downarrow}$. For blue trajectories (converging to $\rm{NP}_{\downarrow}$) the initial condition is at $s_{z}=0.9$ and $\langle \hat{a}^\dag\hat a \rangle =0$. For orange trajectories (limit cycle) the initial condition is at $s_{z}=-0.5$ and $\langle \hat{a}^\dag\hat a \rangle=5$. The two initial states are also specified by $s_{x}=-\sqrt{1-s_{z}^2}$ and $s_y=\langle\hat{X}\rangle=\langle\hat{Y}\rangle=0$. Other parameters: $N=1$, $\omega_0=1$, $g=0.669$, $\omega_z=1.5$, and $\lambda=1.11$}
\end{figure}
%%%%%%%%%%%%%%%%%%%%%%%%%%%%%%%%%%%%%%%%%%%%%%%%%%%%%

It is also interesting to consider the case where the $g_t'$ Hopf bifurcation occurs in a bistable region. An example is the `LC' region of Fig.~\ref{fig2}(b1), where we find a coexistence of limit cycles, linked to the SP phase, and the stable fixed point $\rm{NP}_{\downarrow}$. In Fig.~\ref{fig4} we give the example of two initial conditions belonging to different basins of attraction in phase space. We see in panel (a) how the stable fixed point $\rm{NP}_{\downarrow}$ and stable periodic orbit coexist on the Bloch sphere: If the initial value is in the basin of the stable fixed point $\rm{NP}_{\downarrow}$, the spin trajectory (blue line) converges to the $|\downarrow \rangle $ state, located on the bottom of the sphere. Instead, for another initial condition the trajectory evolves to a stable periodic orbit encircling the (unstable) SP fixed point. The difference in the asymptotic dynamics is also apparent in the time dependence of $\langle \hat{a}^\dag\hat a \rangle$, shown in Fig.~\ref{fig4}(b).

%%%%%%%%%%%%%%%%%%%%%%%%%%%%%%%%%%%%%%%%%%%%%%%%%%%%%

\begin{figure}[htpb!]
	\centering
	\includegraphics[width=0.65\textwidth]{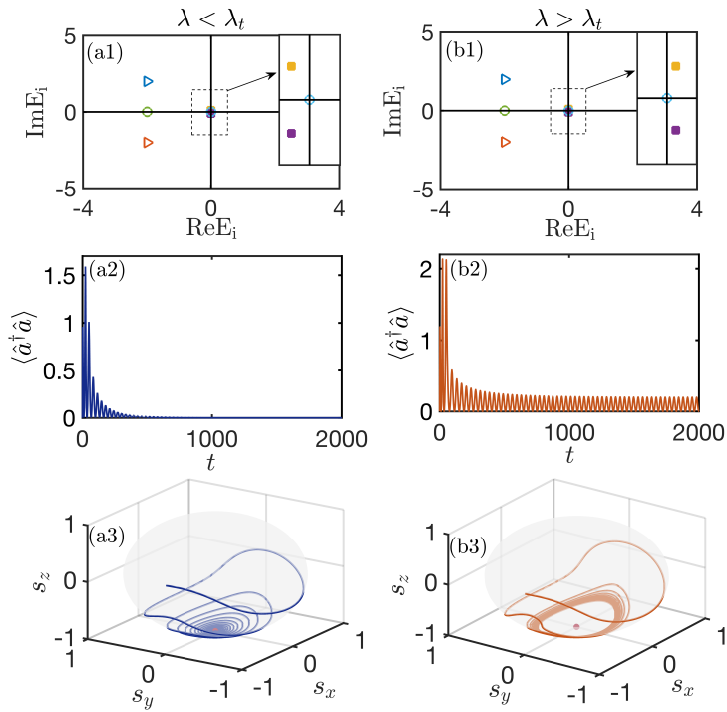}
	\caption{\label{fig5} Hopf bifurcation at $\lambda_t$. The top panels show eigenvalues of the Jacobian matrix of $\rm{NP}_{\downarrow}$, before (a1) and after (b1) bifurcation. (a2) and (a3): Evolution of photon number and trajectories on the Bloch sphere before bifurcation. (b2) and (b3): Evolution after bifurcation. The units of time are determined by $\omega_0=1$. We have used $\lambda=1$ for the left panels and $\lambda=1.1$ for the right panels. The initial condition is $s_{z}=0.2$, $s_{y}=0$, $s_{x}=-\sqrt{1-s_{z}^2}$, and $\langle \hat{a}^\dag\hat a \rangle=\langle \hat{X} \rangle=\langle \hat{Y} \rangle=0$. Other parameters are $N=10$, $g=0.63$, and $\omega_z=1.5$.}
\end{figure}
%%%%%%%%%%%%%%%%%%%%%%%%%%%%%%%%%%%%%%%%%%%%%%%%%%%%%

The second Hopf bifurcation occurs at $\lambda_t$, i.e., coincides with the pole-flip transition and is indicated as thick black lines in Fig.~\ref{fig2}. More precisely, for $\lambda>\lambda_t$ the $\rm{NP}_{\downarrow}$ becomes a limit cycle in phase space, which now coexists with a stable $\rm{NP}_{\uparrow}$ fixed point. Furthermore, we see in Fig.~\ref{fig2} that the $g_t'$ lines continue to the regime $\lambda>\lambda_t$, where a complex coexistence of stable fixed points, different types of limit cycles, and chaotic dynamics occurs. These dynamical features are not shown in the phase diagrams of Fig.~\ref{fig2}, which for $\lambda>\lambda_t$ only consider stable fixed points. While the interplay between different types of dynamics will be discussed in the following section, here we present in Fig.~\ref{fig5} a detailed example of the Hopf bifurcation of $\rm{NP}_{\downarrow}$. In particular, panels (a1) and (b1) show the variation of eigenvalues near the bifurcation. At $\lambda_t$, two conjugate eigenvalues cross the imaginary axis and acquire a positive real part. Before the bifurcation, as shown in panels (a2) and (a3), the system evolves to the stable fixed point $\rm{NP}_{\downarrow}$ with zero photon number. After the bifurcation, a small persistent oscillation in the photon number arises, see panel (b2). At the same time, a periodic orbit forms on the Bloch sphere near the fixed point $\rm{NP}_{\downarrow}$, shown in panel (b3).

%%%%%%%%%%%%%%%%%%%%%%%%%%%%%%%%%%%%%%%%%%%%%%%%%%%%%
\section{ Chaotic dynamics }\label{sec_chaos}
%%%%%%%%%%%%%%%%%%%%%%%%%%%%%%%%%%%%%%%%%%%%%%%%%%%%%

%%%%%%%%%%%%%%%%%%%%%%%%%%%%%%%%%%%%%%%%%%%%%%%%%%%%%
\begin{figure}
\centering
\includegraphics[width=0.65\textwidth]{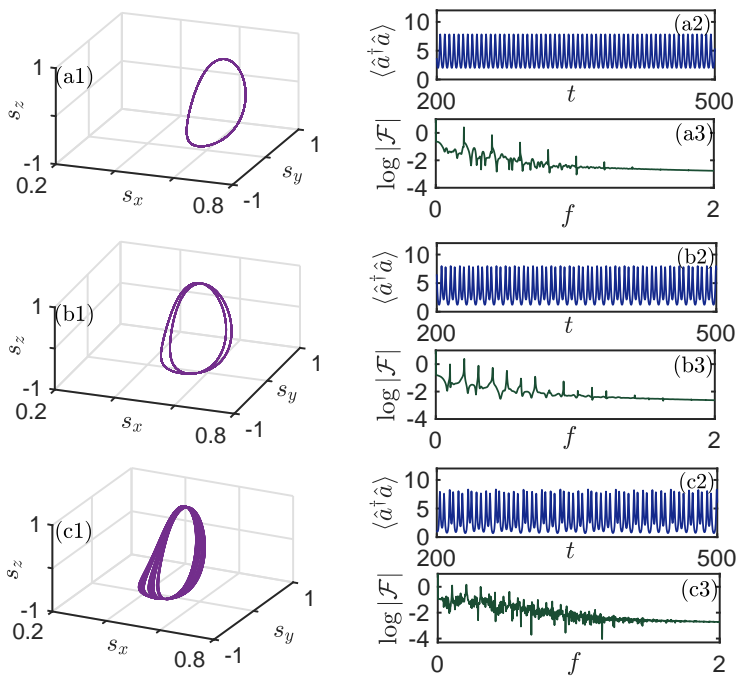}
	\caption{\label{fig6} Evolution from limit cycle to chaotic motion. In panels (a1)-(a3) we use $g=0.9$ (taking $\omega_0$=1). The system evolves on a periodic orbit, forming a single loop in phase space. In panels (b1)-(b3) we use $g=1$. After a period-doubling bifurcation, the system evolves on a periodic orbit with two loops in phase space. In panels (c1)-(c3) we use $g=1.1$. A cascade of period-doubling bifurcations has lead to the formation of a chaotic attractor. $|\mathcal{F}|$ is the power spectral density of $\langle \hat a^\dag \hat a \rangle$. Other parameters: $N=10,\omega_z=1.5,\lambda=1.4$. The initial state has  $s_{z}=-0.99,s_{y}=0$, and $s_{x}=\sqrt{1-s_{z}^2}$.}
\end{figure}
%%%%%%%%%%%%%%%%%%%%%%%%%%%%%%%%%%%%%%%%%%%%%%%%%%%%%

Besides the coexistence of multiple fixed points and limit cycles, chaos emerges in the regime $\lambda>\lambda_t$. The transition to a chaotic trajectory is exemplified in Fig.~\ref{fig6}, where for small coupling strength $g$ the system is in a limit cycle originating from the fixed point $\rm{SP}$. Stable oscillations with equal period and constant amplitude are found in the photon number, see panel (a2). At the same time, the qubit trajectory in panel (a1) is a periodic orbit featuring a single saddle loop. In panel (a3), we have computed the power spectral density (PSD) of $\langle \hat{a}^\dag \hat{a} \rangle$, which reflects the characteristics of system oscillations and is a common tool in the study of chaos. As seen, the PSD shows several discrete sidebands. At larger coupling strength, a period-doubling bifurcation occurs, seen in Fig.~\ref{fig6}(b1). Now the qubit trajectory on the Bloch sphere bifurcates to a new periodic orbit with two saddle loops. The photon number dynamics in (b2) shows periodic oscillations with multiple amplitudes, and the photon number spectrum (b3) has more harmonic peaks. Still, the cyclic motion gives discrete lines in the PSD. By further increasing the coupling strength $g$, chaos eventually appears. The photon number dynamics of panel (c2) shows irregular oscillations with varying amplitude, giving rise to a continuous spectrum. In phase space, the qubit trajectory becomes a complex orbit formed by an infinite family of loops. This example shows that, in our system, chaos emerges from a cascade of period-doubling bifurcations, which is a typical route to chaos.

\begin{figure}
	\centering
	\includegraphics[width=0.55\textwidth]{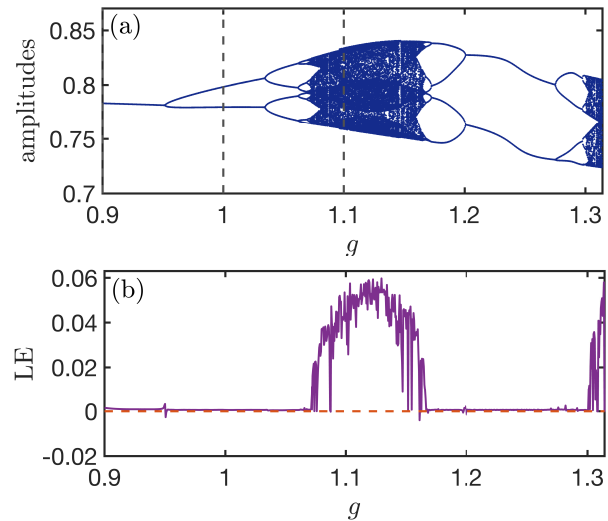}
	\caption{\label{fig7} (a): Bifurcation diagram. The dashed lines mark the period-doubling and chaotic trajectories presented in Fig.~\ref{fig6} (whle the first periodic trajectory is at $g=0.9$). (b): Corresponding values of the Lyapunov exponent (LE). The dashed line denotes the zero value. Other parameters are as in Fig.~\ref{fig6}.}
\end{figure}

The appearance of chaos can be explored in detail through a bifurcation diagram, see Fig.~\ref{fig7}(a). There, we plot the evolution with $g$ of the stationary points in the photon number oscillations. Periodic motion leads to a finite number of amplitudes, while chaotic motion has an infinite number of amplitudes. In the figure, we see that oscillations at small $g$ have a single amplitude, which successively bifurcates to $2^n$ amplitudes ($n=1,2\ldots$). This process finally leads to a dense set of points in a certain range of $g$. The bifurcation diagram is in good agreement with the variation of the Lyapunov exponent $\lambda_{LE}$ with coupling strength, shown in Fig.~\ref{fig7}(b). In the regime with a finite number of amplitudes $\lambda_{LE}$ is very close to zero (dashed line), consistent with periodic oscillations. In the regime with infinite amplitudes, positive values of $\lambda_{LE}$ confirm the occurrence of chaos. 

%%%%%%%%%%%%%%%%%%%%%%%%%%%%%%%%%%%%%%%%%%%%%%%%%%%%%
%%%%%%%%%%%%%%%%%%%%%%%%%%%%%%%%%%%%%%%%%%%%%%%%%%%%%

%%%%%%%%%%%%%%%%%%%%%%%%%%%%%%%%%%%%%%%%%%%%%%%%%%%%%
\subsection{Chaotic regions in the phase diagram}
%%%%%%%%%%%%%%%%%%%%%%%%%%%%%%%%%%%%%%%%%%%%%%%%%%%%%

%%%%%%%%%%%%%%%%%%%%%%%%%%%%%%%%%%%%%%%%%%%%%%%%%%%%%
\begin{figure}
	\centering
	\includegraphics[width=0.65\textwidth]{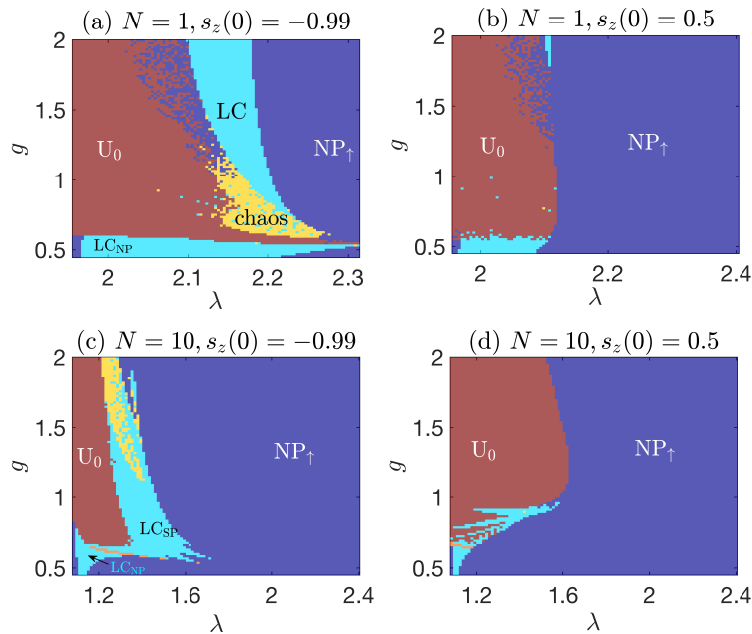}
	\caption{\label{fig8} Phase diagrams at $\lambda>\lambda_t$ for $N=1$ (top) and $N=10$ (bottom). The left and right panels consider two initial states, with $s_z=-0.99$ and 0.5 respectively. Different colors denote various types of dynamics, labeled as in Table~\ref{table1}. Furthermore, we indicate with $\rm{LC}_{\rm{NP}}$ ($\rm{LC}_{\rm{SP}}$) limit cycles around the fixed point $\rm{NP}_{\downarrow}$ ($\rm{SP}$). Periodic motion in the ultra-strong coupling regime of panel (a) is characterized by large oscillations covering both unstable fixed points, thus is labeled as $\rm{LC}$. Chaotic motion occurs in the yellow areas. In panels (c) and (d), the narrow orange strips within the LC regions indicate a $\rm{SP}$ phase. We used $\omega_0=1$, $\omega_z=1.5$, and $s_{x}=\sqrt{1-s_{z}^2},s_{y}=0$ for the initial state.
}
\end{figure}
%%%%%%%%%%%%%%%%%%%%%%%%%%%%%%%%%%%%%%%%%%%%%%%%%%%%%

Accounting for the presence of chaotic dynamics, we can now give a more complete picture of the phase diagram at $\lambda > \lambda_t$. In general, for a given choice of parameters, multiple types of motion coexist in phase space and the system initial values have a significant influence on dynamics. Determining such coexistence for general parameters is a quite involved problem. Therefore, we choose to present the phase diagrams in Fig.~\ref{fig8} by fixing two representative initial states. We use different colors to denote various types of motion: The brown area corresponds to dynamics towards the localized point $\rm{U}_0$; the blue area represents the evolution towards the stable fixed point $\rm{NP}_{\uparrow}$; the light blue area is a phase with periodic oscillations ($\rm{LC}$), and the yellow area indicates chaotic motion. As seen, we may get a completely different phase diagram by changing the initial state. The coexistence of phases is implied by a comparison of corresponding phase diagrams, e.g., panels (c) and (d). 

The left panels of Fig.~\ref{fig8} are for an initial state which is far away from the stable fixed point $\rm{NP}_{\uparrow}$. Close to $\lambda_t$ we find at smaller values of $g$ a region of periodic motion, formed by the fixed point $\rm{NP}_{\downarrow}$ through Hopf bifurcation. This area is labeled by $\rm{LC}_{\rm{NP}}$. Instead, for $\lambda \gtrsim \lambda_t$ and larger $g$, the $\rm{U}_0$ phase appears. The $\rm{NP}_{\uparrow}$ phase dominates the right part of the phase diagrams, with large values of $\lambda$. Periodic oscillations and chaotic motion appear between the areas of $\rm{U}_0$ and $\rm{NP}_{\uparrow}$. In this case, the periodic motion originates from the fixed point $\rm{SP}$, thus we label it as $\rm{LC}_{\rm{SP}}$ in Fig.~\ref{fig8}(c). In this panel, the two types of oscillatory phases are separated by stable dynamics, converging to the fixed point $\rm{SP}$ (orange area). 

The number of qubits $N$ has a marked influence on these phase diagrams. In panel (a), where $N=1$, the chaotic regime occurs at relatively small values of $g$ and large anisotropy $\lambda$. Instead, in panel (c) (where $N=10$) we find that chaos is favored for a larger coupling strength. Besides, in panel (c) the phase $U_0$ appears in a much smaller range of parameters, while the area of fixed points $\rm{NP}_{\uparrow}$ is obviously larger than the area of localized fixed points $\rm{U}_0$ and other types of dynamics. We also note that some periodic windows occur in the chaotic phase, which can be seen in the phase diagram (a) and (c) as blue dots within the yellow region. The occurrence of small windows of regular motion can also be seen in Fig.~\ref{fig7}.

Finally, the right panels of Fig.~\ref{fig8} are for an initial state close to the stable fixed point $\rm{NP}_{\uparrow}$. Now the blue area (where the system converges to $\rm{NP}_{\uparrow}$) occupies a larger portion of the phase diagram, while other phases have shrunk. Especially, periodic motion and chaos have almost disappeared in panels (b) and (d).

%%%%%%%%%%%%%%%%%%%%%%%%%%%%%%%%%%%%%%%%%%%%%%%%%%%%%
\subsection{ Collision of chaotic attractors }
%%%%%%%%%%%%%%%%%%%%%%%%%%%%%%%%%%%%%%%%%%%%%%%%%%%%%

%%%%%%%%%%%%%%%%%%%%%%%%%%%%%%%%%%%%%%%%%%%%%%%%%%%%%
\begin{figure}
	\centering
	\includegraphics[width=0.65\textwidth]{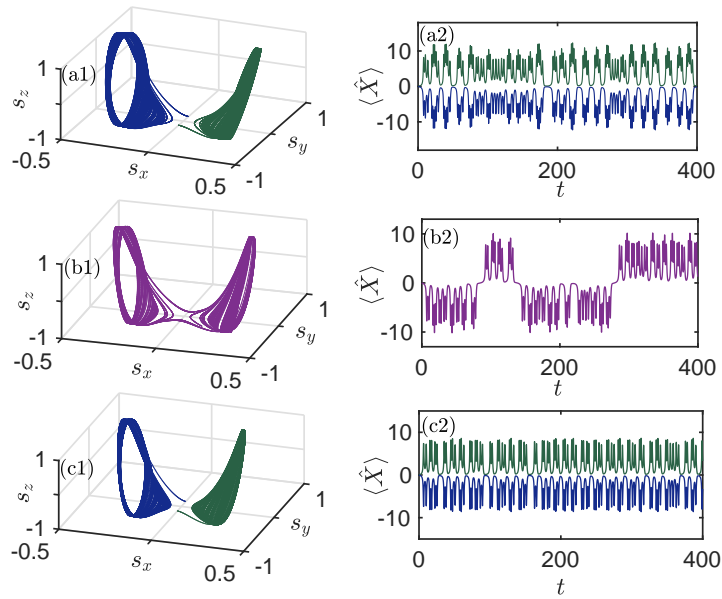}
	\caption{ \label{fig9} Collision and splitting of a pair of symmetric chaotic attractors. (a1)-(a2): At $g=1.8$ there are two isolated chaotic attractors. (b1)-(b2): At $g=2$  the two attractors have merged into a single large attractor. (c1)-(c2): At $g=2.2$ the single large attractor has split again into two isolated chaotic attractors. Here $\lambda=1.25$ and the other parameters are as in Fig.~\ref{fig8}(c).}
\end{figure}

%%%%%%%%%%%%%%%%%%%%%%%%%%%%%%%%%%%%%%%%%%%%%%%%%%%%%

Similar to the one-photon model \cite{stitely2020nonlinear}, we also find the interesting phenomenon shown in Fig.~\ref{fig9}, where two chaotic attractors merge into a single large attractor. As shown in panel (c1), the large attractor can split again into two separated attractors. Usually, when considering symmetric initial conditions ($\pm s_{xi},\pm s_{yi},s_{zi}$), the system evolves in separate parts of the phase space without making transitions between the two regions. This behavior is shown in Fig.~\ref{fig9}(a1) where, after a long time evolution, the two trajectories form two spatially symmetric chaotic attractors. However, by increasing the coupling strength one may observe a dramatic change in system dynamics, shown in panels (b1) and (b2). Now the system makes frequent transitions between the two previously disconnected attractors, thus symmetric initial conditions give rise to an identical large attractor. In this regime, the evolution of the photon field has frequent jumps between different spaces with positive and negative quadrature components $\langle \hat{X}\rangle $, see panel (b2). The collision reveals a distinct route of formation for chaotic attractors. By further increasing the coupling strength, the large attractor splits again into two isolated small attractors, shown in Fig.~\ref{fig9}(c1). The sequence of collision and fragmentation of attractors will continue at larger values of $g$, until the boundary of the chaotic region is reached.

%%%%%%%%%%%%%%%%%%%%%%%%%%%%%%%%%%%%%%%%%%%%%%%%%%%%%
\section{Conclusion}\label{sec_con}
%%%%%%%%%%%%%%%%%%%%%%%%%%%%%%%%%%%%%%%%%%%%%%%%%%%%%

In this article, we examined the nonlinear dynamics of the open two-photon quantum Dicke model in the mean-field approximation. As expected, the mean-field treatment becomes accurate in the limit of a large number $N$ of qubits. By allowing unbalanced rotating and counter-rotating coupling strengths, and taking into account decay of the bosonic field, a rich dynamical behavior is found as a function of system parameters. Several phase diagrams are presented, showing a complex interplay of normal and `superradiant'-like fixed points, as well as localized phases, limit cycles, and chaotic dynamics. 

At variance with the one-photon Dicke model, here the `superradiant'-like phase only displays macroscopic occupation of the atomic degrees of freedom, while the cavity remains in a Gaussian state with finite squeezing and number of photons. A divergence in photon number is found at a critical boundary corresponding to the `spectral collapse' of the closed system. Furthermore, we find various coexistence phases where multiple dynamical behaviors are allowed. Here, the long-time dynamics reflects the segmentation of phase space into different basins of attraction. Especially interesting is the appearance of chaos from a cascade of period-doubling bifurcations. In this chaotic regime, we have highlighted a distinct mechanism of formation of the chaotic attractor through collision and splitting of symmetric lobes.

The two-photon Dicke model finds a natural realization in chains of trapped ions under bicromatic drive \cite{felicetti2015spectral,puebla2017protected,lv2018quantum}. Previous proposals focused on $\lambda=1$, but the anisotropic parameter can be controlled  in a simple way through the relative strength of the two laser drives. Quantum superconducting circuits leading to a similar two-photon interaction, $\propto (\hat a+\hat{a}^\dag )^2\hat{\sigma}_x$, have been discussed in \cite{felicetti2018two,felicetti2018ultrastrong}, and it should be possible to construct alternative schemes realizing Eq.~(\ref{eqH1}). For a given setup, terms neglected in the effective Hamiltonian might have an important effect in certain cases \cite{PhysRevA.98.053819}. In particular, the full model is necessary to describe the long-time behavior in the $\rm{U}_0$ phase.

While here we restricted ourselves to the semiclassical limit, it would be interesting to address how these nonlinear phenomena are reflected by quantum features beyond the mean-field approach. In particular, the occurrence of limit cycles and chaos might be probed through spectral properties of the stationary state and the Liouvillian \cite{PhysRevLett.61.1899,PhysRevLett.123.254101,PhysRevX.10.021019,PhysRevB.101.214302,li2021spectral}, rather than directly from the time evolution. The exponential growth of the out-of-time-ordered correlator (OTOC) is another interesting measure of quantum chaos, recently applied to the closed Dicke model \cite{PhysRevLett.122.024101}. Furthermore, we have supposed here that the dominant decoherence mechanism is from cavity decay. The role of qubit relaxation and dephasing would considerably enlarge the parameter space and possibly induce even richer dynamics, besides having practical relevance. It would be also interesting to explore possible links between the cyclic dynamics we have described and time-crystals in dissipative settings~\cite{iemini2018boundary,gong2018discrete,tucker2018shattered,shammah2018open,heugel2019classical,zhu2019dicke,lledo2019driven,lazarides2020time,waqas2020kinetic,seibold2020dissipative,kongkhambut2022observation,piccitto2021symmetries}. In general, our work provides insight into the rich nonlinear dynamics of the two-photon Dicke model, which may stimulate further investigation of classical and quantum chaos.

%%%%%%%%%%%%%%%%%%%%%%%%%%%%%%%%%%%%%%%%%%%%%%%%%%%%%
\section{Acknowledgments}
%%%%%%%%%%%%%%%%%%%%%%%%%%%%%%%%%%%%%%%%%%%%%%%%%%%%%

R.F. acknowledges partial financial support from the Google Quantum Research Award. R.F. research has been conducted within the framework of the Trieste Institute for Theoretical Quantum Technologies (TQT). S.C. acknowledges support from the National Science Association Funds (Grant No. U1930402) and NSFC (Grants No. 11974040 and No. 12150610464).

%%%%%%%%%%%%%%%%%%%%%%%%%%%%%%%%%%%%%%%%%%%%%%%%%%%%%
\appendix
\section{Validity of mean-field treatment}\label{app1}

%%%%%%%%%%%%%%%%%%%%%%%%%%%%%%%%%%%%%%%%%%%%%%%%%%%%%
\begin{figure}
	\centering
	\includegraphics[width=0.65\textwidth]{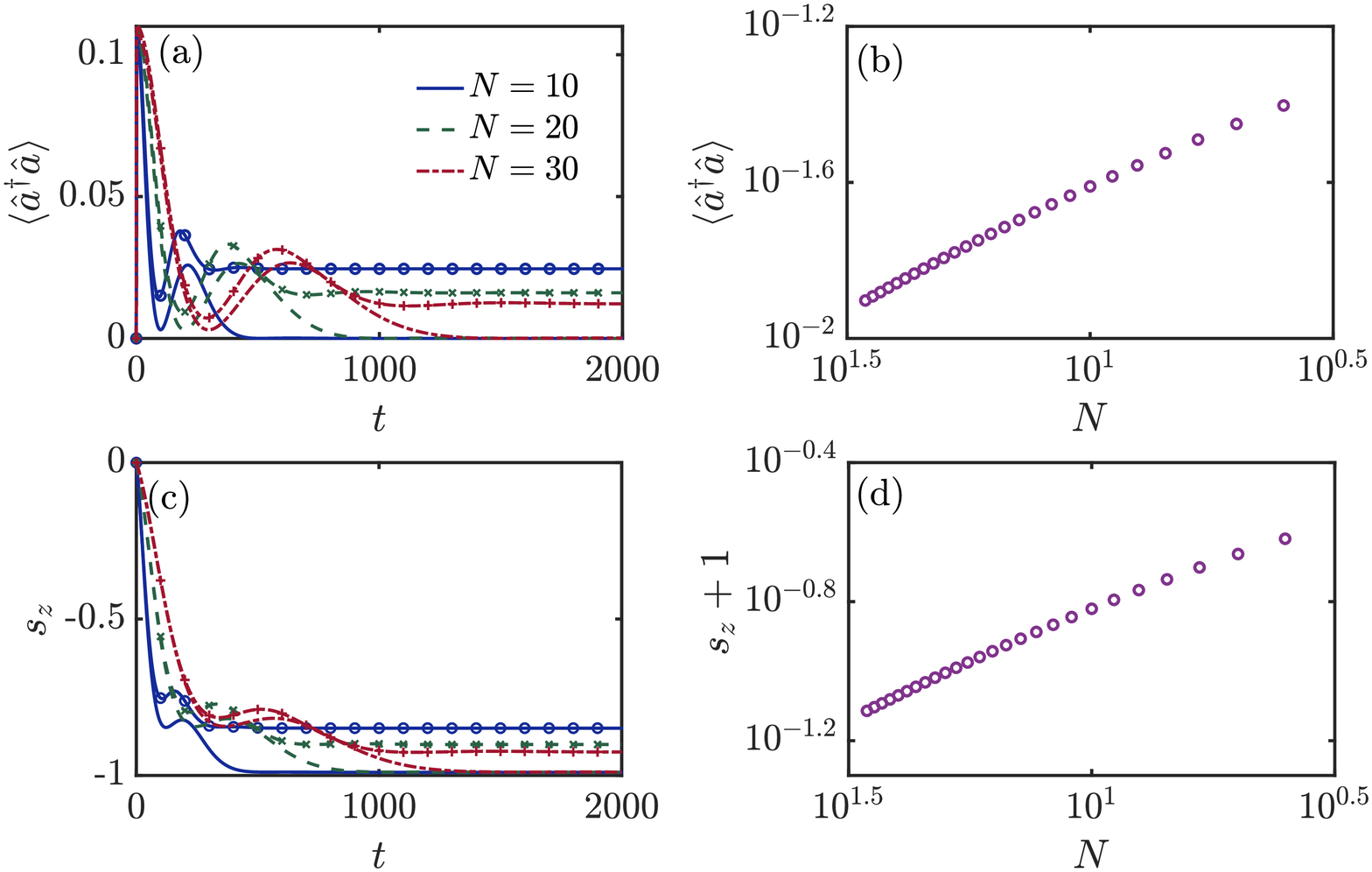}
	\caption{\label{fig_app1} Comparison of mean-field treatment and exact solution in the normal phase. Left panels: full time evolution at different values of $N$, where lines with symbols are calculated by the full master equation~(\ref{eqME}) and lines without symbols are the mean-field approximation, Eqs.~(\ref{eqMFfirst}--\ref{eqMFlast}). Right panels: stationary values as function of $N$, obtained from the master equation, showing that $\langle \hat{a}^\dag\hat{a} \rangle \sim \mathcal{O}(1/N)$ and $s_z \simeq -1+ \mathcal{O}(1/N)$. We used $\omega_0=1$, $g=0.4$, $\omega_z=0.2$, and $\lambda=0.5$. }
\end{figure}
%%%%%%%%%%%%%%%%%%%%%%%%%%%%%%%%%%%%%%%%%%%%%%%%%%%%%

The mean-field approximation is well-known to describe accurately the large-$N$ limit of the one-photon Dicke model. In the two-photon model, however, the cavity does not need to approach a classical state. Here only the atomic variables have a macroscopic population while a non-zero photon number originates from squeezed vacuum fluctuations~\cite{garbe2017superradiant}. In particular, $\langle \hat{a}^\dag \hat{a}\rangle$ does not scale extensively with system size. 

Despite this difference, the $\hat{J}_{x,y,z}$ operators of a large collective spin approach the classical limit, and this is sufficient to justify the mean-field approximation. In particular, we can approximately factorize correlations with the bosonic field (e.g., $\langle \hat{a}^\dag\hat{a} J_z\rangle \simeq \langle \hat{a}^\dag\hat{a}\rangle \langle J_z \rangle $), which is the basic assumption leading to Eqs.~(\ref{eqMFfirst}--\ref{eqMFlast}). The purpose of this Appendix is to test explicitly this property, by comparison to the exact dynamics obtained from the Lindblad master equation~Eq.~(\ref{eqME}).

We first consider in Fig.~\ref{fig_app1} the normal phase regime. In the left panels, (a) and (c), the full time evolution is shown, starting from the same eigenstate of $\hat{J}_x$. We see that the initial agreement between master equation (ME)~\cite{johansson2012qutip,JOHANSSON20131234} and mean-field treatment (MF) holds for longer times when increasing $N$. The ME evolution (solid curves) yields a finite saturation value for $\langle \hat{a}^\dag\hat{a}\rangle $. However, panel (b) shows that $\langle \hat{a}^\dag\hat{a}\rangle  \to 0 $ at large $N$, in agreement with the MF prediction. Similarly, panel (d) shows that $s_z \to -1$ in the stationary state of the ME.

%%%%%%%%%%%%%%%%%%%%%%%%%%%%%%%%%%%%%%%%%%%%%%%%%%%%%
\begin{figure}
	\centering
	\includegraphics[width=0.65\textwidth]{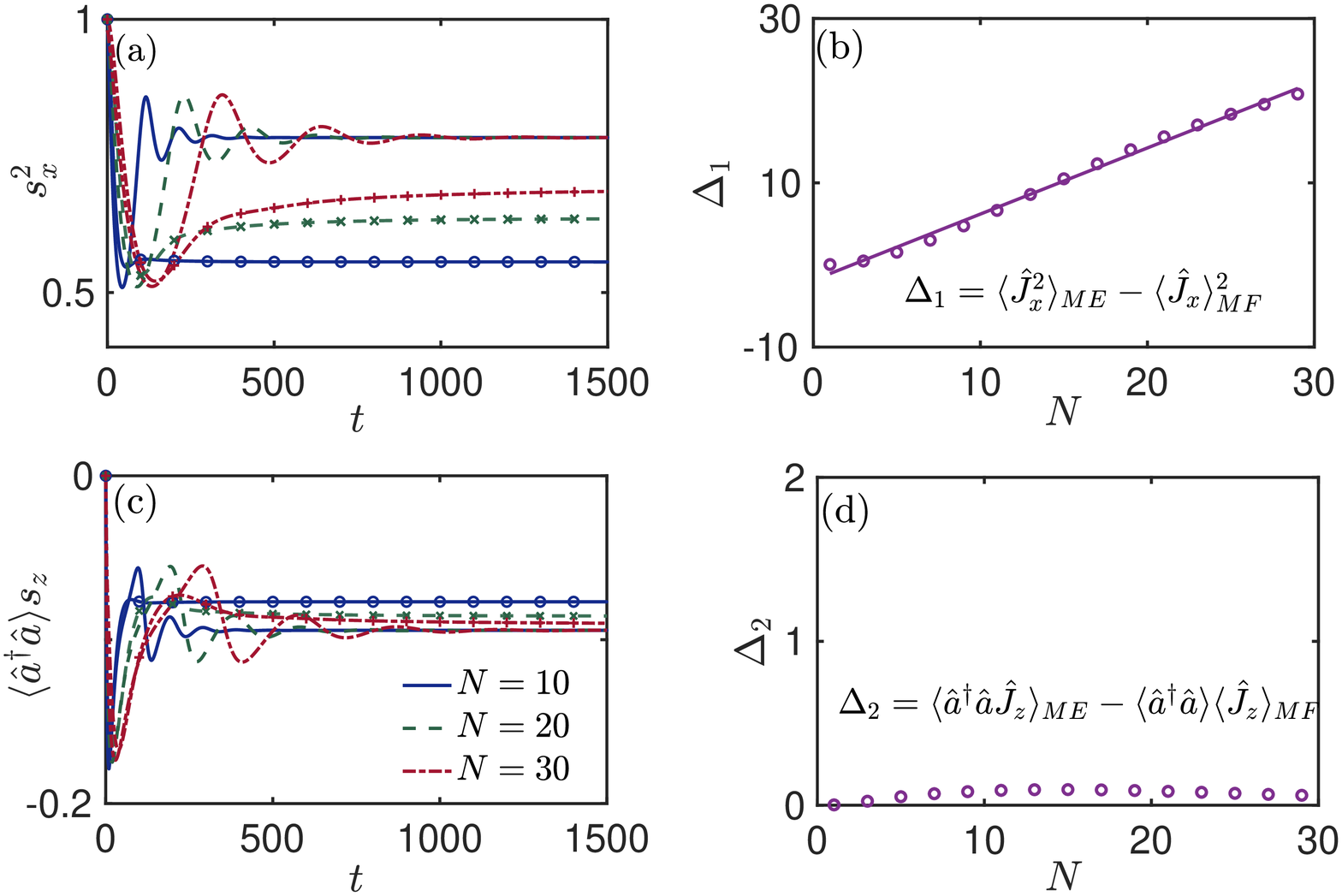}
	\caption{\label{fig_app2}  Comparison of mean-field treatment and exact solution in the `superradiant'-like phase. The curves with symbols in panel (a) and (c) are obtained from the master equation~(\ref{eqME}) while curves without symbols are the mean-field approximation. More precisely, in panel (a) and (c) we use the master equation to compute $4\langle \hat{J}_x^2\rangle/N^2$ and $2\langle \hat{a}^{\dagger}\hat{a}\hat{J}_z\rangle/N$, respectively.  In the right panels, the circles are stationary values calculated from the master equation. The solid line in panel (b) is a linear fit. The parameters are $\omega_0=1$, $g=0.7$, $\omega_z=0.2$, and $\lambda=0.5$.}
\end{figure}
%%%%%%%%%%%%%%%%%%%%%%%%%%%%%%%%%%%%%%%%%%%%%%%%%%%%%

Similar comparisons are shown in Fig.~\ref{fig_app2} for the `superradiant'-like phase. In particular, panel (b) shows that $\langle \hat{J}^2_x\rangle_{ME}-\langle \hat{J}_x\rangle_{MF}^2 \sim \mathcal{O}(N)$. Therefore, the mean-field prediction $\langle \hat{J}^2_x\rangle/N^2 \simeq s^2_x$ is accurate in the limit of large $N$. Panel (d) confirms that the decoupling of $\langle  \hat{a}^\dag\hat{a} \hat{J}_z \rangle$ become justified at large $N$, as the difference $\langle  \hat{a}^\dag\hat{a} \hat{J}_z \rangle_{ME}-\langle  \hat{a}^\dag\hat{a} \rangle \langle \hat{J}_z\rangle_{MF} \sim \mathcal{O}(1)$ is of subleading order if compared to $\langle \hat{J}_z\rangle \sim \mathcal{O}(N)$.

Even if corrections to mean-field follow the expected scaling with $N$, Figs.~\ref{fig_app1} and \ref{fig_app2} show that they remain significant for relatively large system size. At finihe $N$, mean field-theory becomes very accurate when the SP states approach $\rm{U}_0$, due to the macroscopic occupation of the cavity [see Eq.~(\ref{eqSP}), where $s_z \to 0$]. On the other hand, the presence of limit cycles and chaos can be only reflected by the transient quantum dynamics, as the master equation always has a well-defined stationary state. Thus, these regimes are challenging to identify from the quantum evolution with moderate $N$.

%%%%%%%%%%%%%%%%%%%%%%%%%%%%%%%%%%%%%%%%%%%%%%%%%%%%%%%%%%%%
\section{ Coexistence of stable fixed points }\label{sec_multi}
%%%%%%%%%%%%%%%%%%%%%%%%%%%%%%%%%%%%%%%%%%%%%%%%%%%%%%%%%%%%

%%%%%%%%%%%%%%%%%%%%%%%%%%%%%%%%%%%%%%%%%%%%%%%%%%%%%%%%%%%%
\begin{figure}
	\centering
	\includegraphics[width=0.65\textwidth]{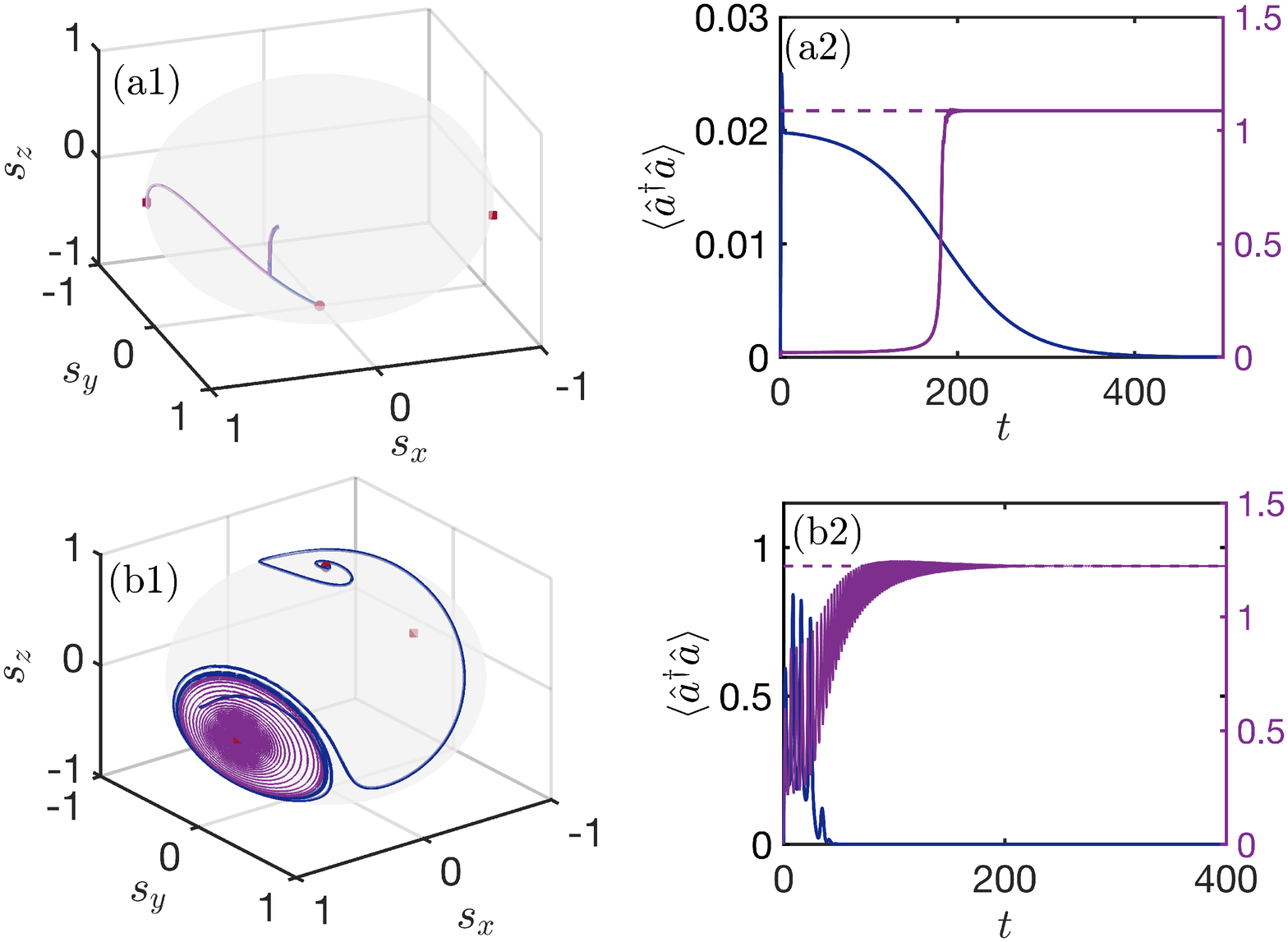}
	\caption{\label{fig_appB1} Evolution in the bistable regimes $\rm{B}_{\downarrow}$ and $\rm{B}_{\uparrow}$. Here we assume $N=1$, $\omega_z=0.2$ (taking $\omega_0=1$), and an initial vacuum state of the cavity. In panels (a1) and (a2) we choose $g=0.85,\lambda=0.5$, which are in the $\rm{B}_{\downarrow}$ (green) area of Fig.~\ref{fig2}(a1). The two initial states have $s_{z}=-0.69$ and $s_{z}=-0.68$ (while $s_x=0$) for the blue and purple lines, respectively. In panels (b1) and (b2) we choose $g=0.45,\lambda=1.8$, which are in the $\rm{B}_{\uparrow}$ (orange) area of Fig.~\ref{fig2}(a1). The two initial states have $s_{z}=0.09$ and $s_{z}=0.08$ (while $s_y=0$) for the blue and purple lines, respectively.}
\end{figure}
%%%%%%%%%%%%%%%%%%%%%%%%%%%%%%%%%%%%%%%%%%%%%%%%%%%%%%%%%%%%

As discussed in the main text, there are regions of parameters where two different types of stable fixed points coexist. The coexistence of ($\rm{NP}_{\downarrow},\rm{SP}$) is indicated as $\rm{B}_{\downarrow}$ and occurs for $\lambda<\lambda_t$. Instead, for $\lambda>\lambda_t$ a bistable phase $\rm{B}_{\uparrow}$ is found. In these phases, as illustrated in Fig.~\ref{fig_appB1}, the asymptotic dynamics is sensitive to the initial condition. In panel (a1) we show two trajectories on the Bloch sphere (left panels) which start at nearby points but eventually diverge from each other, to approach either the $\rm{NP}_{\downarrow}$ fixed point or the `superradiant'-like one. Accordingly, the photon number in panel (a2) either approaches zero (blue line) or reaches a nonzero stable value (solid purple line). The dashed purple line is the steady-state photon number of $\rm{SP}$, given by Eq.~(\ref{eqSP}). Similar behavior can be found in the bistable regime $\rm{B}_{\uparrow}$, whose dynamics of the qubit and photon number are respectively shown in panels (b1) and (b2). 

%%%%%%%%%%%%%%%%%%%%%%%%%%%%%%%%%%%%%%%%%%%%%%%%%%%%%%%%%%%%
\begin{figure}
	\centering
	\includegraphics[width=0.65\textwidth]{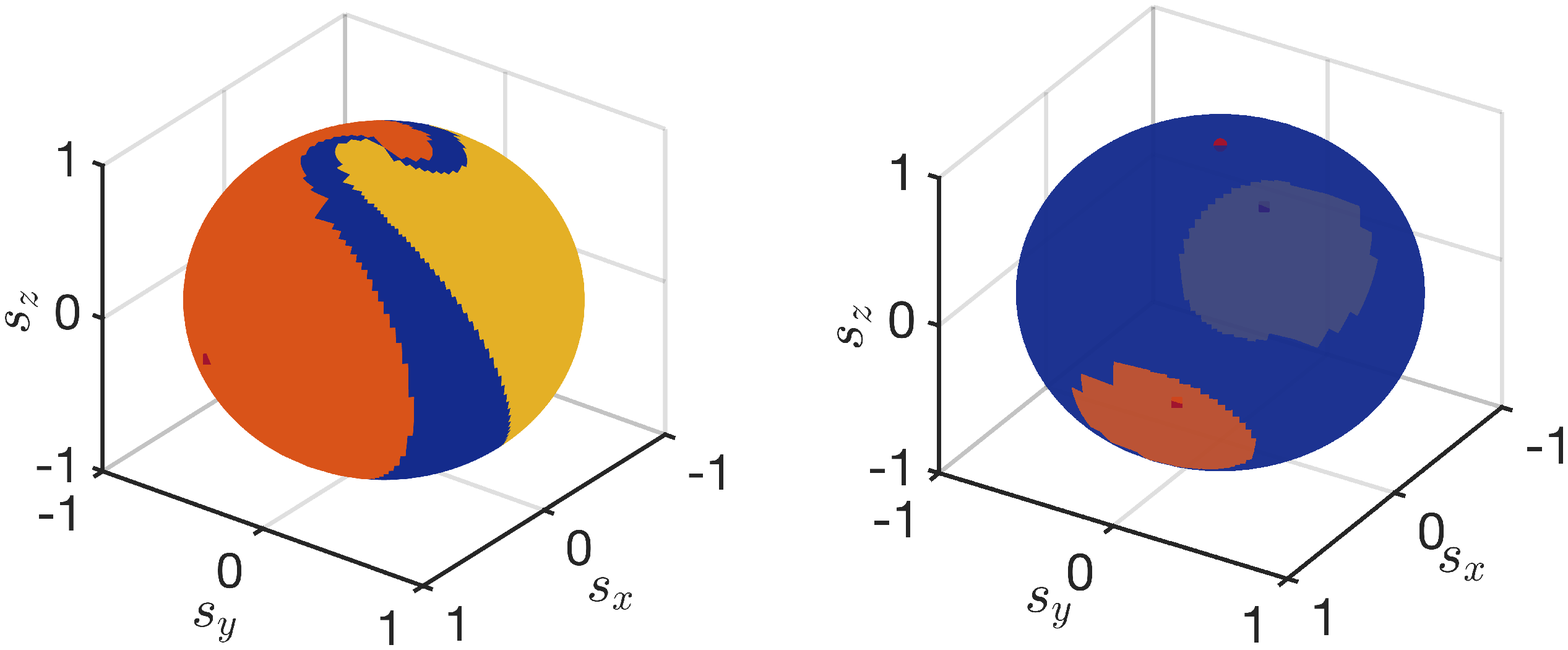}
	\caption{\label{fig_appB2} Basins of attraction on the Bloch sphere, taking the bosonic mode in the vacuum state at the initial time. In the left panel, except for the initial condition of the qubit, the parameters are as in the top panels of Fig.~\ref{fig_appB1} ($\rm{B}_{\downarrow}$ phase). Yellow and orange areas are basins of attraction for $\rm{SP}$, while the blue area is the basin of $\rm{NP}_{\downarrow}$. Right panel: parameters are chosen as in the lower panels of Fig.~\ref{fig_appB1} ($\rm{B}_{\uparrow}$ phase). Here the blue area is the basin of $\rm{NP}_{\uparrow}$. }
\end{figure}
%%%%%%%%%%%%%%%%%%%%%%%%%%%%%%%%%%%%%%%%%%%%%%%%%%%%%%%%%%%%

As each stable fixed point has its basin of attraction, the phase space is divided into distinct regions: Initial states starting from a given basin will all converge to the same stable condition. For definiteness, in Fig.~\ref{fig_appB2} we have chosen  initial values $\langle \hat{a}^\dag \hat{a}\rangle =\langle \hat{X}\rangle = \langle \hat{Y}\rangle=0$ for the bosonic mode (i.e., the vacuum state) and $N=1$. For this choice, in Fig.~\ref{fig_appB2} we have represented graphically the basins of attraction on the Bloch sphere. While the behavior illustrated by Figs.~\ref{fig_appB1} and~\ref{fig_appB2} is generic, we see that the extension and shape of the basins of attraction is sensitive to system parameters. The area of the $\rm{SP}$ basin in the $\rm{B}_{\downarrow}$ example (left side of Fig.~\ref{fig_appB2}) is much larger than basin of $\rm{SP}$ in the $\rm{B}_{\uparrow}$ example (right panel). For the bistable regime $\rm{B}_{\uparrow}$ of Fig.~\ref{fig_appB2}, the phase space is primarily occupied by the $\rm{NP}_{\uparrow}$ basin of attraction.

\begin{figure}
	\centering
	\includegraphics[width=0.65\textwidth]{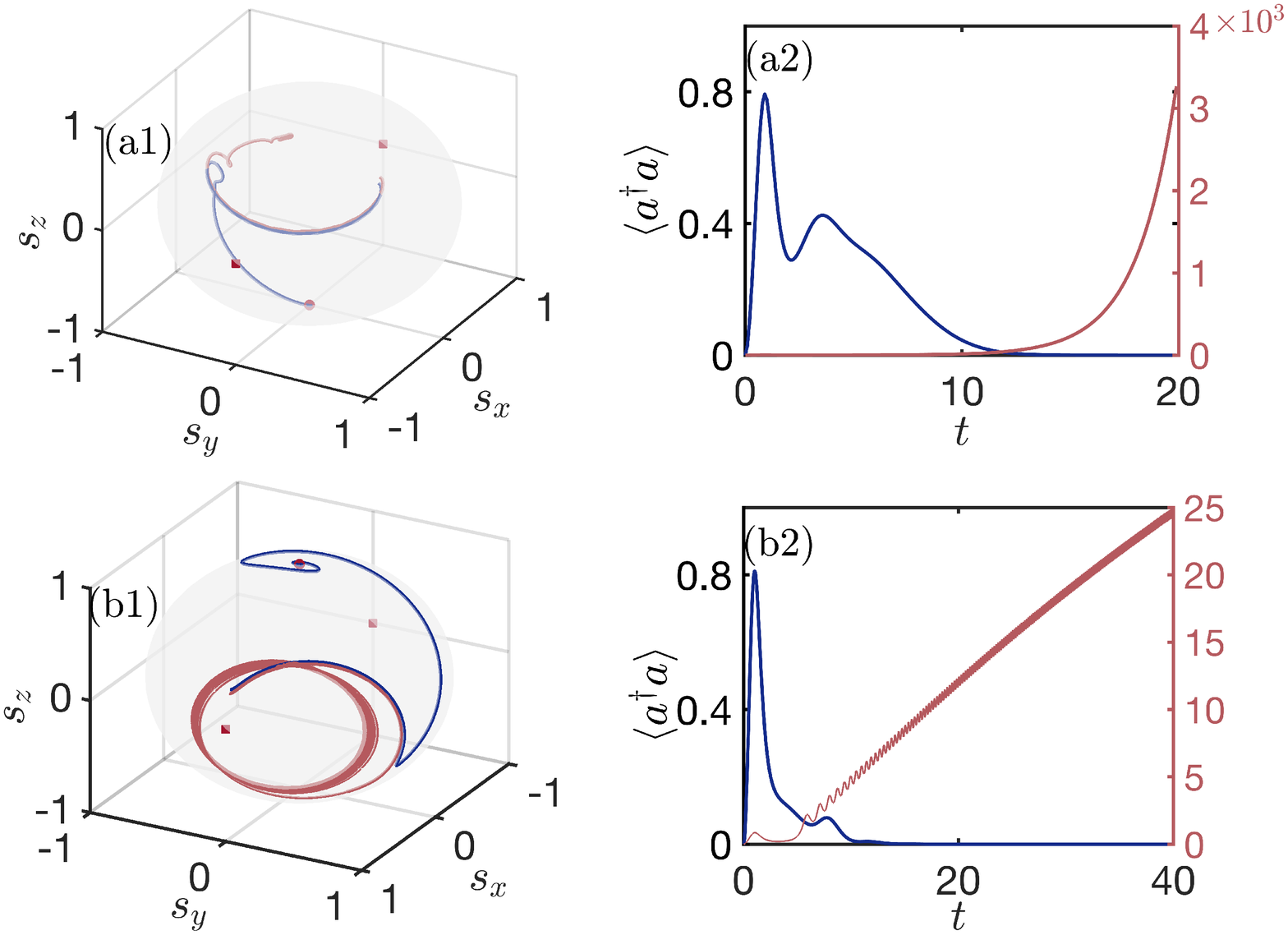}
	\caption{\label{fig_appB3} Evolution in the coexistence phases $\rm{C}_{\downarrow}$ and $\rm{C}_{\uparrow}$. Here we assume $N=1$, $\omega_z=0.2$ (taking $\omega_0=1$), and an initial vacuum state of the bosonic field. In panels (a1) and (a2) we choose $g=1.3,\lambda=0.5$, which are in the $\rm{C}_{\downarrow}$ area of Fig.~\ref{fig2}(a1). The two initial states have $s_{z}=-0.35$ and $s_{z}=-0.3$ (while $s_x=\sqrt{1-s_z^2}$ and $s_y=0$) for the blue ($\rm{NP}_{\downarrow}$) and red ($\rm{U}_0$) lines, respectively. In panels (b1) and (b2) we choose $g=0.65,\lambda=1.8$, which are in the $\rm{C}_{\uparrow}$ area of Fig.~\ref{fig2}(a1). The two initial states have $s_{z}=0.33$ and $s_{z}=0.3$ for the blue and red lines, respectively.}
\end{figure}

For $g>g_{t1}$ the SP phase turns into $\rm{U}_0$, therefore coexistence phases $\rm{C}_{\uparrow/\downarrow}$ appear. The corresponding behavior for representative trajectories is illustrated in Fig.~\ref{fig_appB3}. The discussion is completely analogous to the $\rm{B}_{\uparrow/\downarrow}$ phases, except that now the photon number shows a divergent time dependence for initial conditions in the $\rm{U}_0$ basin of attraction. For other initial conditions, see the blue curves of Fig.~\ref{fig_appB3}, the system approaches rapidly the normal phase fixed point without displaying any singularity.

%%%%%%%%%%%%%%%%%%%%%%%%%%%%%%%%%%%%%%%%%%%%%%%%%%%%%%%%%%%%
%%%%%%%%%%%%%%%%%%%%%%%%%%%%%%%%%%%%%%%%%%%%%%%%%%%%%%%%%%%%

\section*{References}
\bibliography{chaosTAQRM}

\end{document}